\documentclass{article}

\usepackage{arxiv}

\usepackage[utf8]{inputenc} 
\usepackage[T1]{fontenc}    
\usepackage{hyperref}       
\usepackage{url}            
\usepackage{booktabs}       
\usepackage{amsfonts}       
\usepackage{nicefrac}       
\usepackage{microtype}      
\usepackage{lipsum}
\usepackage{graphicx}

\title{OpenStreetMap data use cases during the early months of the COVID-19 pandemic\thanks{Please use the final version of this article as:\newline Mooney, P., Grinberger, A. Y., Minghini, M., Coetzee, S., Juhász, L., \& Yeboah, G. (2021). OpenStreetMap Data Use Cases During the Early Months of the COVID-19 Pandemic. In: Rajabifard, Foliente \& Paez (Eds.), \textit{COVID-19 Pandemic, Geospatial Information, and Community Resilience – Global Applications and Lessons.} CRC Press (Taylor \& Francis) (pp. 171-185) doi: \href{https://doi.org/10.1201/9781003181590-15}{10.1201/9781003181590-15} }}

\author{
  Peter Mooney\\
  Department of Computer Science\\
  Maynooth University, Ireland\\
  \texttt{peter.mooney@mu.ie} \\
   \And
 A. Yair Grinberger \\
  Department of Geography\\
  Hebrew University of Jerusalem, Israel\\
  \texttt{yair.grinberger@mail.huji.ac.il} \\
 \AND
  Marco Minghini\thanks{The views expressed are purely those of the authors and may not in any circumstances be regarded as stating an official position of the European Commission.} \\
  Digital Economy Unit \\
  European  Commission,  Joint  Research  Centre, Italy \\
  \texttt{marco.minghini@ec.europa.eu} \\
  \And
  Serena Coetzee \\
   Department of Geography, Geoinformatics and Meteorology \\
  University of Pretoria, South Africa \\
  \texttt{serena.coetzee@up.ac.za} \\
  \And
  Levente Juhasz \\
  GIS Center \\
  Florida International University, USA \\
  \texttt{ljuhasz@fiu.edu} \\
    \And
Godwin Yeboah \\
  Institute for Global Sustainable Development\\
  University of Warwick, UK \\
  \texttt{g.yeboah@warwick.ac.uk} 
}

\begin{document}
\maketitle

\section*{Abstract}
\label{section:abstract}

Created by volunteers since 2004, OpenStreetMap (OSM) is a global geographic database available under an open access license and currently used by a multitude of actors worldwide. This chapter describes the role played by OSM during the early months (from January to July 2020) of the ongoing COVID-19 pandemic, which - in contrast to past disasters and epidemics - is a global event impacting both developed and developing countries. A large number of COVID-19-related OSM use cases were collected and grouped into a number of research frameworks which are analyzed separately: dashboards and services simply using OSM as a basemap, applications using raw OSM data, initiatives to collect new OSM data, imports of authoritative data into OSM, and traditional academic research on OSM in the COVID-19 response. The wealth of examples provided in the chapter, including an analysis of OSM tile usage in two countries (Italy and China) deeply affected in the earliest months of 2020, prove that OSM has been and still is heavily used to address the COVID-19 crisis, although with types and mechanisms that are often different depending on the affected area or country and the related communities.


\section{Introduction}
\label{section:introduction}

The OpenStreetMap (OSM) project was started in University College London in 2004 and has subsequently grown to be arguably the largest and most popular Volunteered Geographic Information (VGI) and open geographic data project in the world today \cite{OSM_wiki}. The data within the OSM database is completely open and is available under an Open Database License (ODbL). The contributors of geographic data to OSM are predominantly citizens not specifically connected to the professional production or management of geographic data. However, in recent years, governments and commercial companies have become involved in the contribution of data to OSM and the editing and maintenance of the existing data. Mooney and Minghini~\cite{mooney_review_2017} provide an extensive description of the users and uses of OSM data. Many people who encounter OSM mistakenly consider the online maps and associated digital cartographic products from OSM as the entire OSM project. This is incorrect. OSM is primarily a very large spatial database of geographic data. Online maps and other services such as routing or information services are derived products and could not exist without the underlying OSM database. The data model used within the OSM database is simple to understand but is powerful enough to prove capable of expressing the complex geographical relationships and topologies encountered in real world environments such as road networks, commercial and industrial settings, landuse features such as lakes and rivers, and residential buildings.

The data model expresses three object types: nodes (or points), polygons and polylines (collectively called ways). Relations are a logical object expressing a collection of these objects to represent complex compound geographic features such as railway stations, airports, transport routes, etc. Every object (except those nodes making up ways) must have at least one descriptive attribute associated with it. These attributes are called tags and are stored as key-value pairs. Very detailed guidance on the available tags, acceptable values for specific keys, and usage examples are provided on the Map Features page in the OSM Wiki \cite{OSM_wiki_MF} and often within software. However, the application and use of tags on objects in OSM is not strictly enforced and follows a folksonomy approach allowing contributors to choose tags as they see appropriate. This approach has led to many criticisms of the quality of OSM data over the years \cite{Seto_2020}. One unique aspect of the OSM database is the ability for anyone to access the entire contribution and editing history of the data within the database. This allows researchers to study the evolution of the OSM data in specific areas, study contribution patterns over time, analyse how the OSM database grows with influence from external events such as natural or humanitarian disasters. 

There are a number of methods which support contribution or insertion of geographic data into the OSM database. These methods are supplemented by a myriad of software tools and services available with the entire OSM ecosystem. Field survey, implying a physical knowledge of the area under survey, using GPS tools, cameras and other software is supported widely. Social events called ‘mapping parties’ often involve people meeting up at group events to undertake field mapping of a specific area (\cite{mooney_observations_2015,Brovelli2018}). However, the most commonly used approach is remote mapping using web-based interfaces such as the popular iD editor allowing contributors to remotely map an area by digitizing data on top of satellite imagery. Contributors using this approach are urged to have some knowledge of the area being mapped and at minimum consult the extensive documentation and guidance on the OSM wiki on how to map properly. ‘Mapathons' are popular events where people, even located in different parts of the world, meet virtually and do remote mapping on the same area. Finally, the software-automated import of existing geographic datasets and databases is also possible and has been used widely in OSM to import datasets such as road networks and buildings. Automated imports are complex database operations and those undertaking such imports are strongly encouraged to seek the approval of the local OSM community in the geographic area of the import before proceeding. Imports should also be clearly documented in the OSM wiki. 

Previous to the COVID-19 pandemic OSM had been used in many humanitarian and environmental disaster situations where access to up-to-date and accurate geographic data was immediately required and remote mapping and field mapping exercises could quickly generate geographic data for an area if none existed. This chapter will analyse and understand how OSM has been used during the early months of the COVID-19 pandemic. By early months we are referring to the period between January 2020 and July 2020. In January 2020 the World Health Organization (WHO) announced the COVID‐19 epidemic, a public health emergency of international concern. The coronavirus disease 2019 (COVID‐19) has subsequently become a global pandemic and has imposed unprecedented change all over the world in how we interact as humans, in our working practices, and how medical professionals carry out their work. Most countries in the world have experienced many COVID-19 related deaths and high rates of COVID-19 spread amongst their populations. This chapter will be a strong and defined contribution to the knowledge on how VGI initiatives such as OSM respond and are used or accessed during a global crisis such as the COVID-19 pandemic. The novel aspect of this work is that this critical assessment is being delivered during this unfolding and unprecedented event rather than from an a posteriori position. Furthermore, we believe that this work will produce knowledge about humanitarian mapping in a context never studied before - a global event with significant impacts in developed regions.

The remainder of the chapter is organized as follows. In Section~\ref{section:background} we provide a discussion of background and related work. Section~\ref{section:methodology} provides an overview of the methodology and research employed in this work. In Section~\ref{section:use-of-osm-data} we discuss the use of OSM data in the COVID-19 response, while section~\ref{section:collectionOSM} discusses the collection of new OSM data for COVID-19 responses. Section~\ref{section:academic} briefly discusses current academic research with OSM during the COVID-19 response. In section~\ref{section:conclusions} we make some conclusions and outline some future work. 


\section{Background and Related Work}
\label{section:background}

The potential of OSM for supporting humanitarian efforts during crisis situations was noticed as early as 2010 after a magnitude 7.0 earthquake struck Haiti. Volunteers across the world supported humanitarian efforts through mapping activities across multiple platforms, including OSM contributors who produced much data using available aerial imagery within only a few weeks~\cite{zook_volunteered_2010}. These efforts have led to the formation of the Humanitarian OpenStreetMap Team (HOT), an international charitable organization which organizes and oversees open mapping in humanitarian contexts~\cite{humanitarian_openstreetmap_2020}. Since then, HOT was involved in initiating and coordinating mapping efforts in multiple cases, including following the 2013 Yolanda Typhoon in the Philippines~\cite{poiani_potential_2016} and the 2015 Nepal Earthquake~\cite{palen_success_2015}. The open source Tasking Manager (TM) software~\cite{humanitarian_openstreetmap_team_hotosmtasking-manager_2020}, developed by HOT and aimed at coordinated collaborative mapping, was also re-used by several communities worldwide, such as the Italian one to coordinate mapping after the 2016 earthquake~\cite{minghini_collaborative_2017}. Today, digital humanitarianism in OSM goes beyond disaster response with the organization of activities that aim at supporting vulnerable communities to increase their resilience~\cite{scholz_volunteered_2018,crowd2map,missing_maps_putting_2020}. Furthermore, the richness of OSM data facilitates further applications, frequently including the development of third party tools that utilize OSM data for creating additional data or carrying geographic analyses. For example, efforts after the 2010 Haitian earthquake also included the application of an Emergency Route Service relying on up-to-date OSM data~\cite{neis_collaborative_2010}; OSM data was also used as a baseline for the Flooded Streets tool used to produce a crowdsourced map of flooded streets in Chennai, India during a 2015 flooding event~\cite{nitin_flooded_2016}.

One issue that is also being considered within this context is the management of health crises and the possible contributions of OSM to managing health crises and monitoring epidemics. Mooney et al.~\cite{mooney_potential_2013}, when discussing the role of VGI in pervasive health applications, identify OSM as a potential ``virtual audit instrument'' describing local environments. Accordingly, much effort is being put into collecting information required for monitoring health-related outcomes, e.g. mapping settlements and buildings to assist malaria prevention in Kenya and Mozambique~\cite{solis_engaging_2018}, health facilities\cite{chan_mapping_2012}, and other critical infrastructure~\cite{herfort_towards_2015}, while other works strive to utilize existing data to mitigate health effects and assess accessibility to medical facilities~\cite{das_location_2014,ferguson_using_2016,lindholm_new_2017,yeboah_godwin_examining_2020}. The response to the 2014-2016 Ebola outbreak in West Africa, which turned from a local response to an extended international effort~\cite{dittus_mass_2017} presents a relevant and unique example. During the breakout, HOT volunteers mapped large portions of the infected regions, providing support for on-the-ground teams of the Médecins Sans Frontières and facilitating the production of epidemiological maps~\cite{lessard-fontaine_supporting_2015,peckham_satellites_2017}. Additionally, an OSM-based navigation service (OSM Automated Navigation Directions - OsmAnd) was used to support data collection activities by locals~\cite{nic_lochlainn_improving_2018}.
The studies surveyed above present crisis relief and disaster response as a multi-dimensional framework. The support of relief efforts may begin with providing reliable basemaps but extends even beyond the production of the information that these maps require into the development of new products and services, based on OSM data, for the benefit of responders and the general population.

In the following sections we use this to survey and analyse the different dimensions through which the OSM community has responded to the COVID-19 crisis. Bearing in mind the cross-boundary effects of the pandemic, we also consider the formation of inter-regional collaborations, as in the case of the Ebola outbreak discussed above.


\section{Methodology and Research Approach}
\label{section:methodology}

\begin{figure}[htb]
\includegraphics[width=1.0\textwidth]{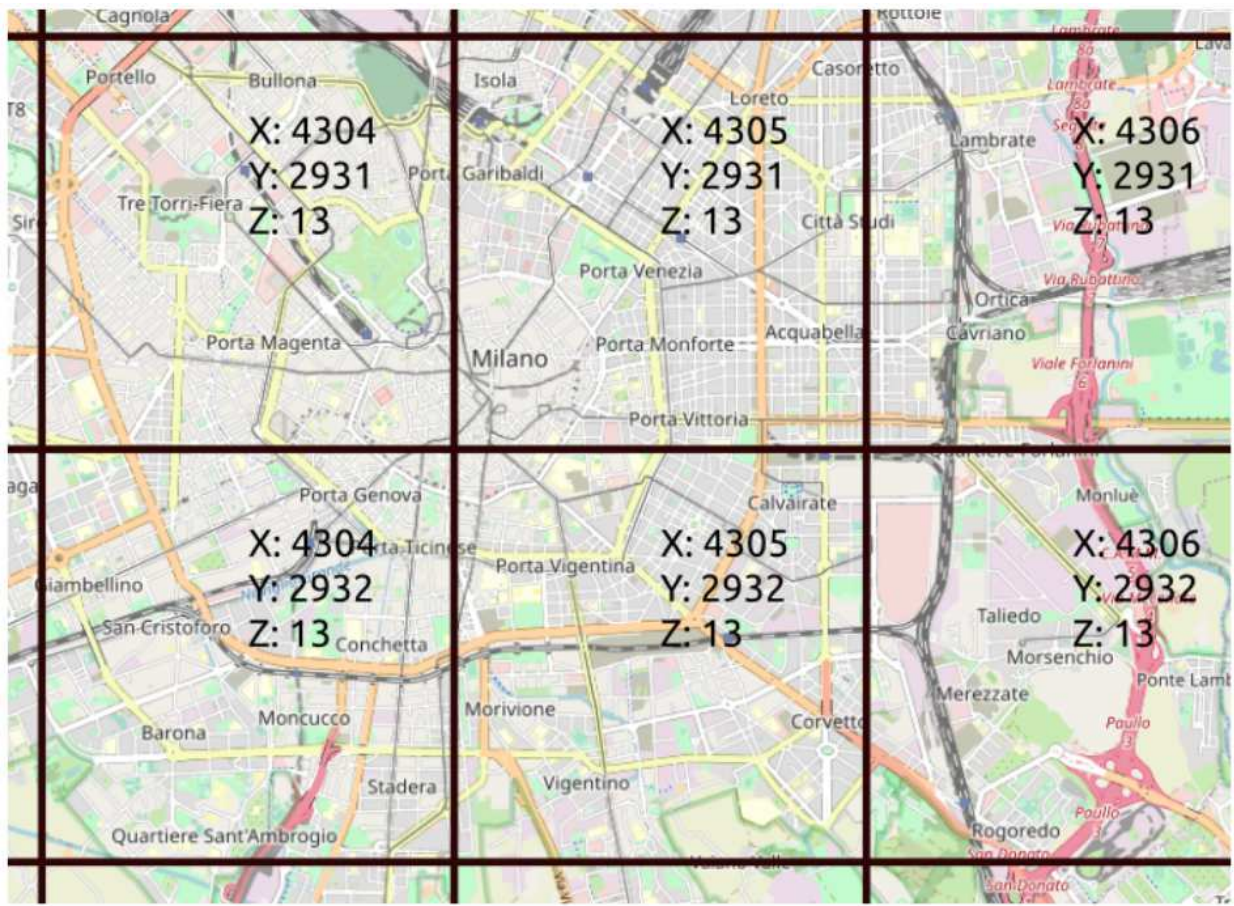}
\caption{Illustration of web map tiles in zoom level 13 in the area of Milan, Italy.}
\label{figure:webtiles}
\end{figure}

To survey and analyse the different dimensions through which the OSM community has responded to the COVID-19 crisis we considered the following \textit{OSM Response Frameworks} (our terminology) as follows:
\begin{itemize}
 \item \textbf{OSM usage as a cartographic basemap in COVID-19 related applications} which can indicate the project’s maturity and ability to compete as an alternative to commercial and authoritative mapping service providers;
 \item \textbf{COVID-19 related applications or services using OSM data} (such as points of interest, road networks, building data such as hospitals or medical facilities);
 \item \textbf{Initiatives or applications aimed at the collection of new OSM data} immediately relevant to the COVID-19 pandemic response or management;
 \item \textbf{COVID-19 influenced imports of authoritative geospatial data into OSM} where there are gaps in the OSM database for a particular country or region; and,
 \item \textbf{Academic research about the role of OSM in the COVID-19 response}. 
\end{itemize}
The proposed methodology for understanding the role played in each of the OSM Response Frameworks is comprised of a number of research tasks which are summarised as follows:
\begin{itemize}
    \item traditional literature review focused on standard academic sources, web searches of social media, and research of available gray literature such as multimedia, reports, presentations and mailing lists;
    \item analysis of OSM map tile access and usage on a global scale, including comparison with the pre-COVID-19 situation to find out whether the pandemic has generated more OSM tile access and usage than pre-COVID-19 situation.
\end{itemize}
While most readers will be familiar with traditional literature reviews, the tiled web map system requires some explanation. The tiled web map system divides the earth into a set of regular tiles corresponding to different zoom levels. Web-based maps usually display geographic information by loading map image tiles (or lately, vector tiles) corresponding to a geographic area and stitching them together to a visually seamless map experience. A description of tiled web maps is found in Juhasz and Hochmair~\cite{juhasz_user_2016}. OSM tiles can be displayed on any map free of charge if adhering to the tile usage policy~\cite{openstreetmap_foundation_operations_working_group_tile_2020}. We extracted OSM tile usage statistics from Planet OSM~\cite{openstreetmap_foundation_planet_2020} for affected areas in Italy (Lombardy) and China (Wuhan) along with control areas within the same country (Sicily and Beijing, respectively) to reveal whether COVID-19 increases tile usage. Our detailed methodology and technical details can be found in Figure~\ref{figure:webtiles}. We choose zoom level 13, in which details correspond to the regional level with one tile covering 23.9 $km^2$ (Figure~\ref{figure:webtiles}) and compare usage between pre-COVID-19 (January 1 - 21) and affected times (February 5 - 25 in China and March 11 - 31 in Italy).

\section{Use of OSM data for COVID-19}
\label{section:use-of-osm-data}

\begin{figure}
\includegraphics[width=1.0\textwidth]{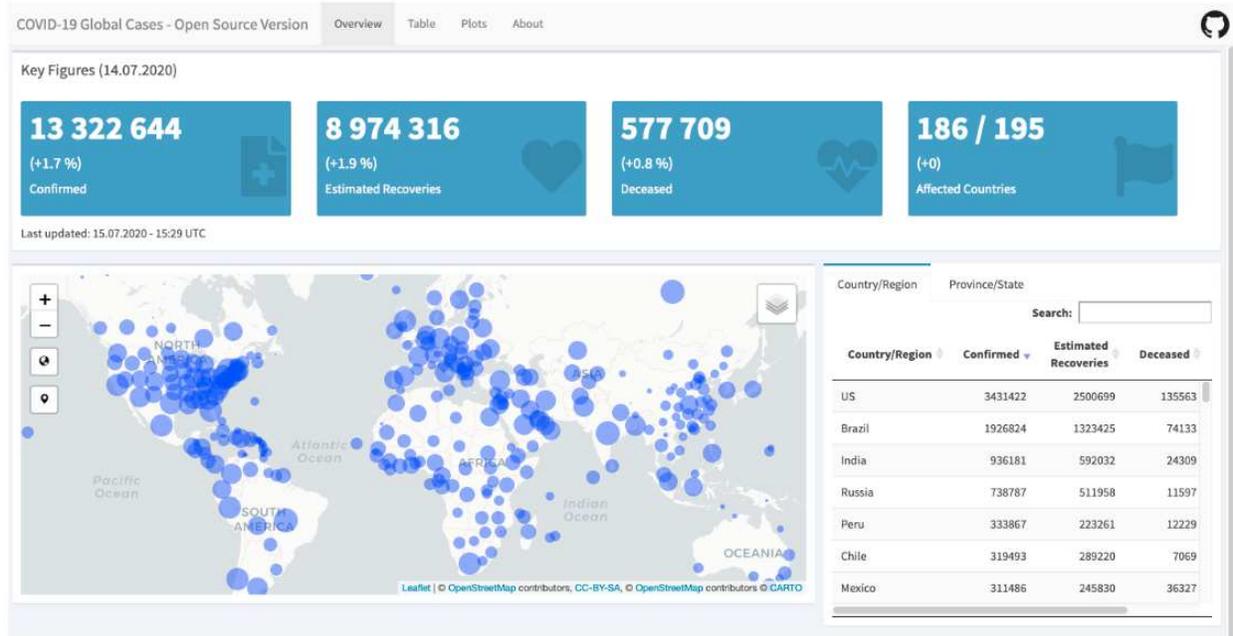}
\caption[OSM COVID-19 web visualizations] {A dashboard focused on global COVID-19 cases. Source:~\cite{schoenenberger_covid-19_2020}.}
\label{figure:COVID-19_cases_FigA}
\end{figure}

\begin{figure}
\includegraphics[width=1.0\textwidth]{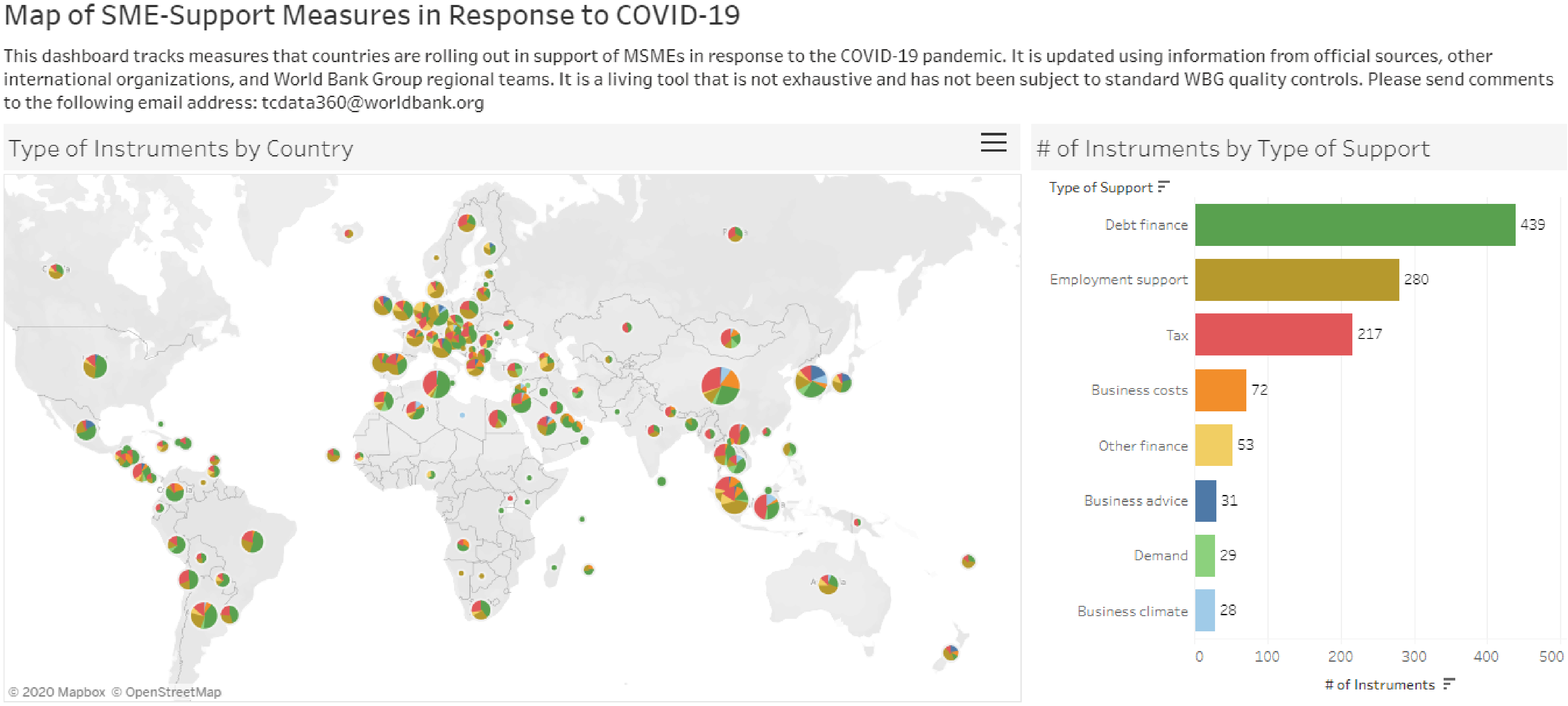}
\caption[World Bank: COVID-19 web visualizations] {World Bank map showing the measures of each country to support small and medium-sized enterprises (SMEs) in response to COVID-19. Source:~\cite{the_world_bank_map_2020}.}
\label{figure:COVID-19_cases_FigB}
\end{figure}

We searched for websites offering web maps and geospatial services related to the COVID-19 pandemic. The bulleted list below (at the end of this Section~\ref{section:use-of-osm-data}) presents a typology of the websites with references to prominent examples. We identified two major types of websites - visualizations of COVID-19 data and geospatial services. The most prominent example of data visualizations are dashboards which simply overlay OSM data with other types of data, i.e. OSM was used as a basemap only (see some examples in Figure~\ref{figure:COVID-19_cases_FigA} and Figure~\ref{figure:COVID-19_cases_FigB}). Online dashboards are typically driven by open data on the pandemic released by governments and/or other organizations (e.g. the popular dataset from the Center for Systems Science and Engineering at Johns Hopkins University~\cite{johns_hopkins_university_center_for_systems_science_and_engineering_jhu_csse_covid-19_2020}) and communicate numbers and statistics, usually updated on a regular basis, through tables, graphs and thematic maps. These dashboards serve to inform experts, decision makers and the general public. Many of the dashboards designed by national governments, research organizations and volunteers use OSM as the basemap. A subtype within this group of websites are websites presenting dynamic visualizations of the spread of the virus. It seems that most of the dashboards featuring OSM basemaps are also realized with open source mapping software and designed by volunteers, and research or not-for-profit organizations and businesses. Dashboards produced by governments (which show the same COVID-19 data) seemed instead to rely more heavily on proprietary technology and basemaps.

\begin{figure}
\includegraphics[width=1.0\textwidth]{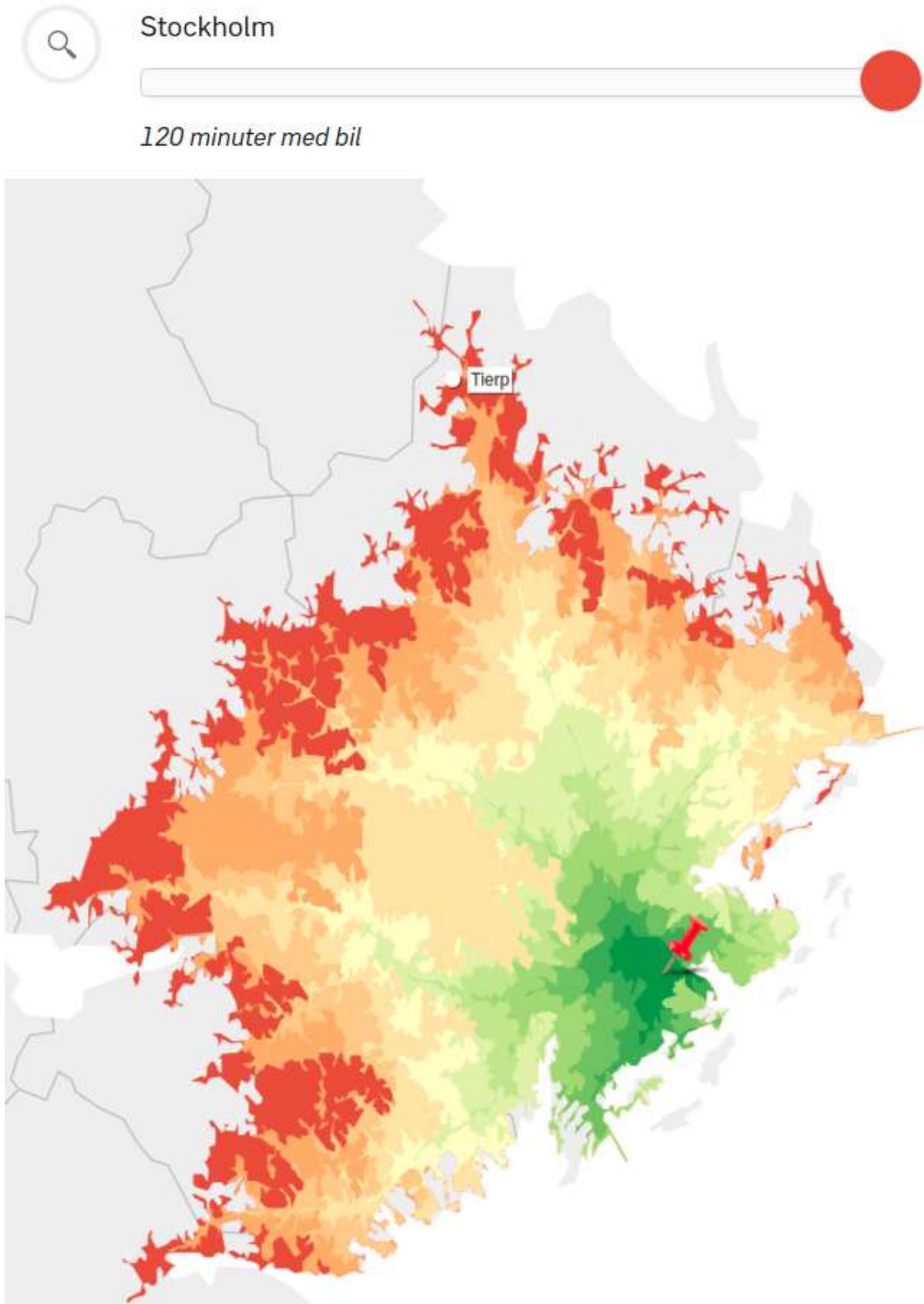}
\caption[Swedish Television’s interface using OSM road network data] {Swedish Television’s interface using OSM road network data and the isochrones function of OpenRouteService for computing areas within 2 hours driving distance from a location in Sweden (example shows the isochrones from Stockholm’s center). Source:~\cite{sveriges_television_ab_sa_2020}.}
\label{figure:sweeden}
\end{figure}

The geospatial services group includes three subtypes. The first provides more types of geospatial information, extending beyond contagion patterns. The information content in these websites is very diverse, ranging from general information on issues such as resilience and support measures for enterprises to practical information, e.g. locations for testing for COVID-19 and on the availability of masks in pharmacies. The second subtype extends this approach (and hence there is some overlap between the two subtypes), allowing users to contribute data themselves, i.e. become ``produsers''~\cite{coleman2009volunteered}. Notice that these are platform-specific data (e.g. where masks are 3-d printed) and not data that are fed back into OSM. The third subtype facilitates more complex spatial queries, e.g. by comparing users’ location history to assess their exposure to COVID-19 or when identifying areas allowed for travel in the vicinity of users’ homes. Swedish TV~\cite{sveriges_television_ab_sa_2020} stands out among these by utilizing the isochrones functionality of OpenRouteService~\cite{heidelberg_institute_for_geoinformation_technology_openrouteservice_2020} (see Figure~\ref{figure:sweeden}, thus extending beyond the ``OSM as basemap'' type of data usage. This is related to the unique restrictions imposed in Sweden which were defined by driving time instead of Euclidean distance, hence requiring more complex spatial querying capabilities.

\paragraph*{\textbf{Typology of websites and services using OSM data, with examples}}

\begin{itemize}

\item \textbf{Website type: COVID-19 data visualizations}

	\begin{itemize}
	\item \textbf{\textit{Dashboards:}} \newline \textit{Examples}:
	Global \cite{johns_hopkins_university_center_for_systems_science_and_engineering_jhu_csse_covid-19_2020,schoenenberger_covid-19_2020,scriby_inc_coronavirus_2020,berliner_morgenpost_corona_2020,university_of_pretoria_and_university_of_neyshabur_covid-19_2020};
	continental/regional \cite{european_commission_joint_research_centre_ecml_2020}; national/local \cite{paolicelli_coronavirus_2020,berliner_morgenpost_corona_2020,the_detroit_news_map_2020,geo-spatialorg_coronavirus_2020,government_of_alberta_covid-19_2020,nichols_maps_2020,observatory_earth_analytics_nigerias_2020};
	\item \textbf{\textit{Dynamic visualizations:}}\newline  \textit{Examples}: 
	the spread of virus infections over time~\cite{healthmap_covid-19_2020}; the phylogeny of SARS-CoV-2 viruses from the ongoing novel coronavirus COVID-19~\cite{nextstrain_team_nextstrain_2020}.
	\end{itemize}

\item \textbf{Website type: COVID-19 related geospatial services}

	\begin{itemize}
	\item \textbf{\textit{Geospatial information services:}} \newline \textit{Examples}:
	measures supporting small to medium-sized enterprises~\cite{the_world_bank_map_2020}; social media posts~\cite{moro_kenya_2020}; change in mobility patterns~\cite{c-19_global_data_science_project_quantifying_2020}; 3-d printing of masks for medical staff~\cite{kaplan_open_source_3d_2020}; locations of clinical trials~\cite{heidelberg_institute_for_geoinforamtion_technology_mapping_2020}; locations of food resources~\cite{city_of_boston_map_2020}; resilience measures for businesses~\cite{grey_county_economic_development_tourism__culture_community_2020} and population~\cite{national_administrative_department_of_statistics_colombia_vulnerabilidad_2020}; locations for testing for COVID-19~\cite{victoria_state_government_getting_2020}; stocks of masks in pharmacies~\cite{kiang_stocks_2020}; queuing time in border crossings~\cite{sixfold_gmbh_truck_2020}; change in air quality during lockdowns~\cite{european_environment_agency_eea_air_2020}; state of public transport~\cite{kaplan_open_source_changes_2020};

	\item \textbf{\textit{Services facilitating (non-OSM) data contribution:}}\newline \textit{Examples}:
	3-d printing of masks for medical staff~\cite{kaplan_open_source_3d_2020}; initiating joint delivery of goods~\cite{kaplan_open_source_want_2020}; queuing times in supermarkets~\cite{filaindiana_filaindiana_2020};

	\item \textbf{\textit{Services performing geospatial queries:}}\newline  \textit{Examples}:
	assessment of exposure to COVID-19 using location history~\cite{israelcoronamap_israelcoronamap_2020,kaplan_open_source_dont_2020,rampazzo_200_2020}; identification of areas allowed for travel from a location, given travel restrictions \cite{kaplan_open_source_how_2020,sveriges_television_ab_sa_2020}.
	\end{itemize}
\end{itemize}

\begin{figure}
\includegraphics[width=1.0\textwidth]{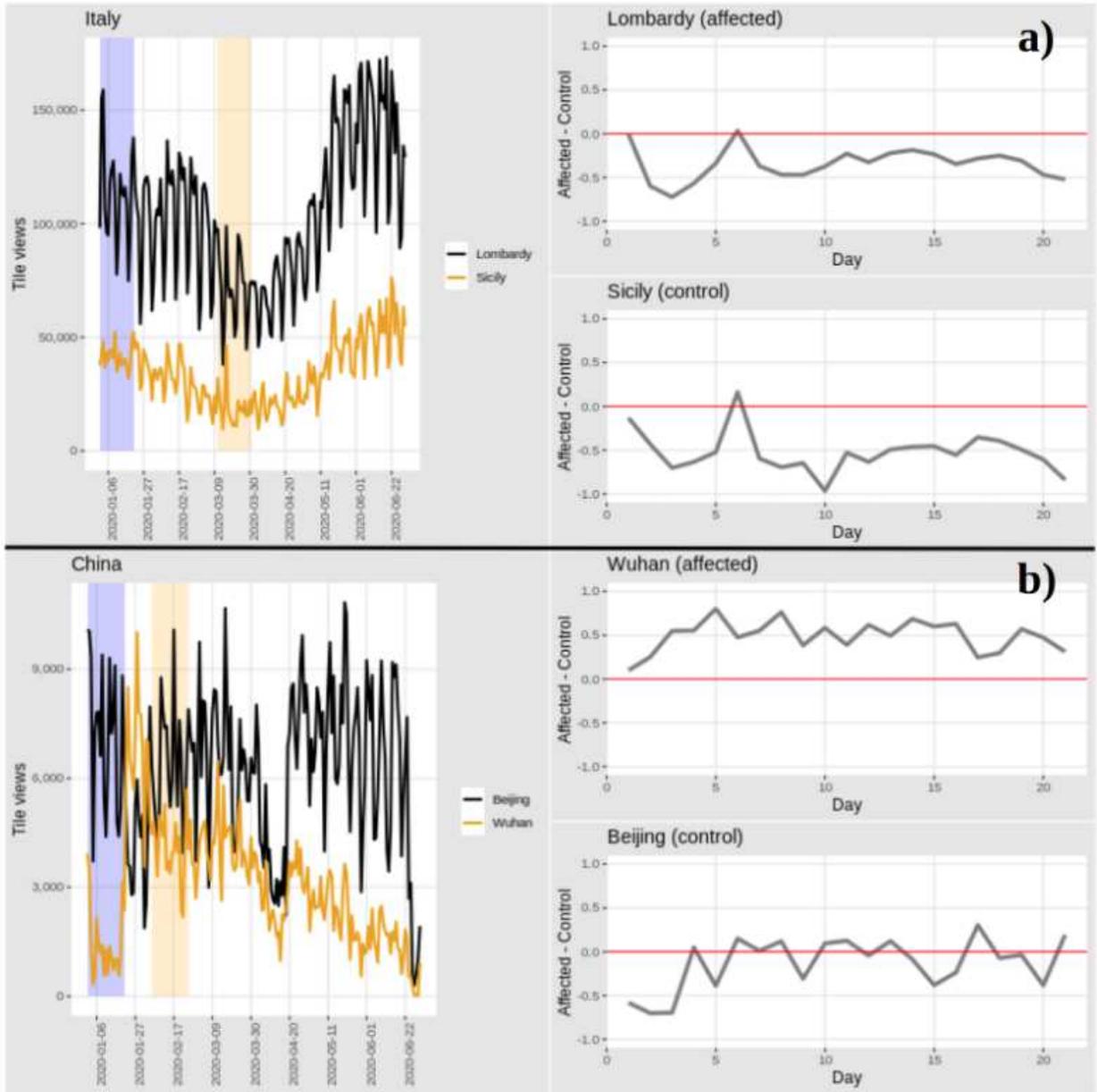}
\caption[OSM Tile usage statistics] {Tile usage statistics for selected areas in Italy (a) and China (b). The left side of figures shows the number of times tiles were loaded. Purple vertical area is the control period where orange vertical area shows a 3-week-period affected by COVID-19. The right side of sub-plots show the difference of normalized tile usage patterns for an area.}
\label{figure:tileusage}
\end{figure}

The dashboards and web maps above make use of a wide range of different OSM basemaps. Many of these are provided by commercial mapping companies such as CARTO~\cite{carto} and Mapbox~\cite{mapbox}. Other websites and services, especially community-developed projects without financial means to afford using commercial providers, rely on the freely available tiles provided by OSM. To assess whether OSM tiles were used in connection with COVID-19 response, we compared tile usage statistics between two greatly affected regions (Lombardy in Italy and Wuhan in China) with their relatively unaffected counterparts within the same countries (Sicily and Beijing). The left panel of Figure~\ref{figure:tileusage}. a-b plots the number of times tiles were loaded for study sites between January 1 and June 30, 2020. The baseline for the comparison was set to a 3-week-long period between January 1 and January 21 (purple, shaded vertical area), which were compared to 3-week-long affected periods (orange, shaded vertical area, February 5 - 25 in China and March 11 - 31 in Italy). Control and affected periods start with a Wednesday and end with a Tuesday 3 weeks later to eliminate the daily temporal trend that is visible in the left side of Figure~\ref{figure:tileusage}. We assume that the seasonal trend is constant across affected areas and their unaffected counterparts within the same country, therefore seasonal patterns were left untreated. Plots in the right panel of Figure~\ref{figure:tileusage} show the difference between normalized tile usage patterns for a region. A value of 1 means that tiles were only loaded during the period affected with COVID-19, and -1 means the opposite. The red horizontal shows that equal numbers of tiles were loaded during the control and affected periods on a given day. The normalized difference is higher in Lombardy and in Wuhan than in Sicily and Beijing respectively, which suggests that areas greatly affected by COVID-19 were viewed more frequently than would be expected under normal conditions as seen in the tile logs. Two paired t-tests were conducted, which confirmed that the increased attention affected areas were experiencing was statistically significant, $t(20) = 5.00$, $p < 0.001$ for Italy and $t(20) = 8.63$, $p < 0.001$ for China.

\section{Collection of OSM data for COVID-19}
\label{section:collectionOSM}

The ongoing COVID-19 pandemic has also seen an unprecedented amount and type of activities to collect new OSM data. Not surprisingly, May 2020 has set the all-time records for the numbers of daily OSM contributors ($7,209$), newly registered OSM users ($6,259$) and newly registered users who contributed data ($1,019$) - all of them on May 14~\cite{openstreetmap_new_2020}. This section summarizes the main nature of such activities and provides, whenever possible, details on the reasons for collecting OSM data and/or the communities or organizations which actually requested those data. However, a key strength of any VGI project is the chance that data can be used by anyone, at any time, and for purposes that might be different and even unknown to the users who originally collected those data~\cite{Mooney_2016}. Proliferation of Artificial Intelligence (AI), which has reached a massive uptake during the COVID-19 crisis \cite{craglia_artificial_2020}, does nothing but reinforce this statement.
OSM data collection to address the COVID-19 pandemic has happened in all the ways described in Section~\ref{section:introduction}. Remote mapping is by far the method by which most data was contributed and it is not surprising that such efforts were led by HOT. At the time of writing (mid-July 2020), the HOT TM lists $183$ projects targeted at COVID-19 emergencies worldwide with aims for the collection of baseline OSM data such as buildings, road networks, land use areas and placenames~\cite{humanitarian_openstreetmap_team_tasking_2020}. These projects mainly address regions in African and South American countries where baseline maps are still not available, with Peru being the most popular one with a total of 84 projects. The OSM contribution records mentioned above were directly attributed to increased activity in HOT’s projects in Peru, Botswana and Central African Republic, with a mapping peak in the Cusco region in Peru~\cite{openstreetmap_new_2020}. A recent tweet from HOT~\cite{humanitarian_openstreetmap_team_hotosm_2020_twitter} reported about more than 10,000 volunteers who have mapped over 1.7 million buildings and over 41,000 km of roads in COVID-19 projects so far. The organizations requesting the activation of these HOT projects, which will use the collected OSM data afterwards, are national or regional governments, health authorities, humanitarian organizations and NGOs. As usual, also during COVID-19 times mapathons (mostly virtual, given the mobility restrictions) have been extensively organized by several organizations worldwide to perform coordinated remote mapping in specific areas, with HOT itself providing tips and suggestions on how to map COVID-specific OSM objects~\cite{humanitarian_openstreetmap_team_volunteer_2020}.

In addition to the HOT TM, another tool that has been widely used during the COVID-19 crisis is \texttt{healthsites.io}~\cite{healthsitesio_building_2020}, which aims to build an open geospatial dataset of every health care facility in the world, allowing to map e.g. hospitals (\texttt{amenity=hospital}), pharmacies (\texttt{amenity=pharmacy}) and doctors (\texttt{amenity=doctors}) and to add tags to the already available ones. This type of mapping clearly requires a personal knowledge of the health facilities to add and therefore it is not a task for remote users like those involved in HOT projects. Similarly in the MapRoulette application for fixing OSM data bugs~\cite{neerhut_improving_2020}, projects were created for improving health-related OSM data, e.g. by adding information about the number of beds in hospitals. Other different types of OSM mapping activities require field surveys to record the locations of specific objects. As an example, in Cape Town (South Africa) communal pit latrines pose a COVID-19 transmission risk, similar to other places frequented by many people, such as public transport, shopping centres or communal water taps. In the Cape Town area (Dunoon), the Western Cape Government used OSM to map at least 900 communal toilets in informal areas, and these were included in their risk analysis and risk management approach~\cite{du_preez_openstreetmap_2020,corburn2020slum,gibson2020novel}.

\begin{figure}
\includegraphics[width=1.0\textwidth]{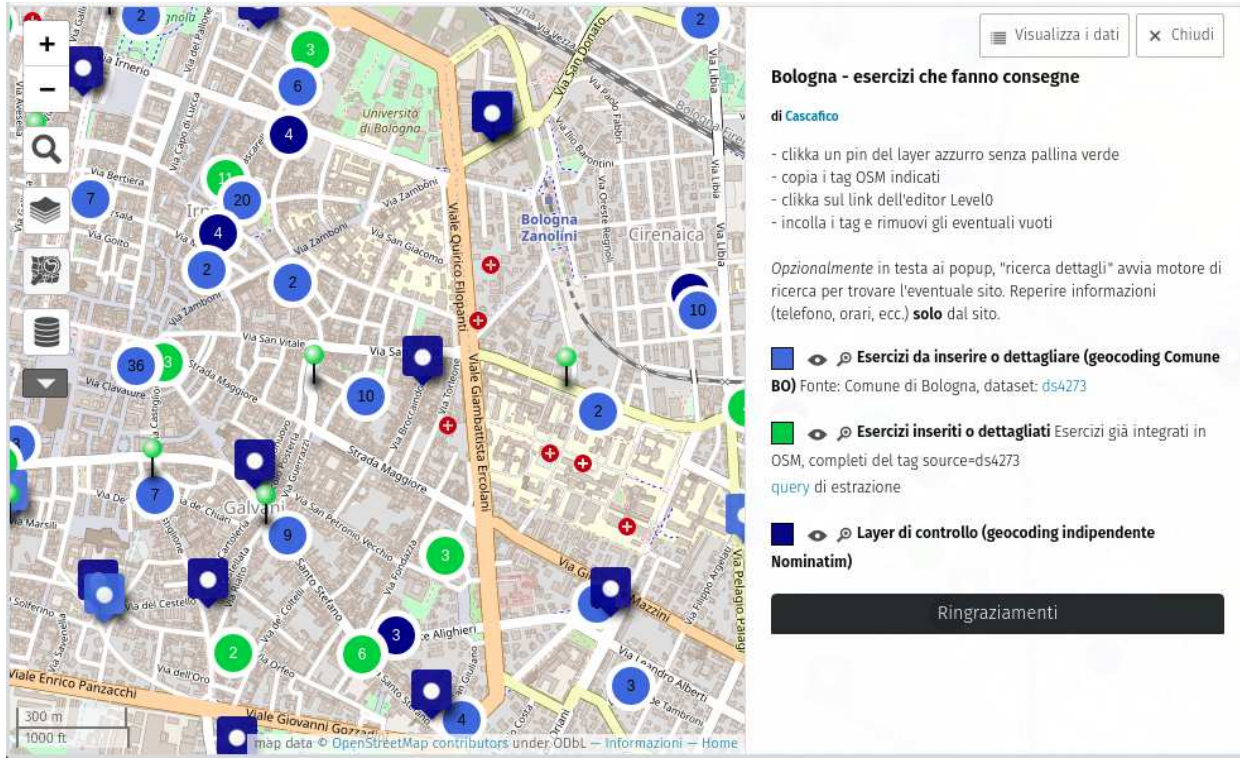}
\caption[OSM import in Bologna, Italy] {uMap project to facilitate the OSM import of commercial activities offering delivery service in the Municipality of Bologna, Italy. The colors of the OSM markers distinguish the activities already imported from those still to be imported. Source:~\cite{italian_openstreetmap_community_umap_2020}.}
\label{figure:bologna_osm}
\end{figure}

The ways to contribute OSM data during the COVID-19 pandemic have been very different in other countries where the baseline cartography was already available and the focus was placed on adding detailed COVID-19 information. For example, the popular application \textit{Ca reste ouvert}~\cite{ca_reste_ouvert_it_2020}, created by the OSM French community and then extended to other countries (including Italy, Germany, Austria and Switzerland), offers a thematic visualization of commercial activities based on whether they are open during the COVID-19 crisis; information on the COVID-19 specific opening hours, takeaway and delivery service are also shown and can be added/edited by users. This explains how dynamically the OSM communities reacted to the emergency by creating new OSM tags such as \texttt{opening\_hours:covid19=*},  \texttt{takeaway:covid19=*}, \texttt{delivery:covid19=*}. Given that such information was either not available elsewhere or made available only in a very fragmented way (e.g. lists of activities that were open or offered takeaway/delivery services were published as plain text on websites of local governments or newspapers), OSM has acted as the only platform to store and offer such data in a structured way. As an exception to the general situation described above, some Italian municipalities provided the datasets of commercial activities offering delivery services on their open data portals under OSM-compatible licenses to allow their import. An example is a dataset from the Municipality of Bologna\cite{comune_di_bologna_coronavirus_2020}, that was imported in OSM by the Italian community through a documented procedure~\cite{italian_openstreetmap_community_importcataloguebo-esercizicommerciali_2020} based on a collaborative uMap project~\cite{italian_openstreetmap_community_umap_2020} (see Figure~\ref{figure:bologna_osm}).

Another key dataset in Italy that was made available for integration in OSM was the dataset of all Italian pharmacies provided by the Italian Ministry of Health. In this case, given that most pharmacies were already available in OSM, the work performed by the community was not a bulk import but a manual integration of the missing information. Finally, given the high use of OSM data in Italy from emergency agencies such as Red Cross, Civil Protection and fire fighters, the Italian community also decided to focus the mapping efforts during COVID-19 times on substantial imports that were prepared or started in the past but not yet completed, e.g. the one for all addresses in the Municipality of Milan~\cite{italian_openstreetmap_community_importcataloguemilan_2020}.


\section{Academic research with OSM during the COVID-19 response}
\label{section:academic}

Although the literature covering COVID-19 is fast changing, there is evidence that OSM is a valuable resource for the scientific community. Published research in the early months of the pandemic mainly appeared in medical and health related outlets, however, at the time of writing, there are several examples of utilizing OSM data and related infrastructures spanning across different scientific disciplines. French-Pardo et al.~\cite{franch-pardo_spatial_2020} reviewed 63 articles on the spatial dimensions of COVID-19 and found that most national and regional web viewers use the ArcGIS Online platform~\cite{arcgis_online}, however, they also noted that some works utilized OSM because it was free. Some studies used a passive approach and utilized OSM data only for display, such as showing aggregate survey results based on neighborhoods extracted from OSM in Israel~\cite{rossman_framework_2020}, or displaying detected hotspots on an OSM basemap~\cite{hohl_daily_2020}. Another study went beyond the basic use of OSM data and extracted social concentration places (e.g. ATMs, bus stops) from OSM to predict mortality trends as part of the first comprehensive study of COVID-19 in Iran~\cite{pourghasemi_spatial_2020}. Qazi et al.~\cite{qazi_geocov19_2020} compiled more than $524$ million COVID-19 tweets and used OSM’s Nominatim service to geocode and reverse geocode toponyms found in tweets~\cite{qazi_geocov19_2020}. This highlights that the ecosystem built around OSM data provides researchers with free to use tools for a number of use cases. Apart from these few examples, a quick literature search on Google Scholar for ``OpenStreetMap'' and ``Covid-19'' keywords yields several early stage research hosted on arxiv, medrxiv, ResearchGate and other preprint publishing services.

If one had to share data representing the home locations of COVID-19 infected people, their location privacy would be infringed. Geographic masks make it possible to share data in a representative way, without risking the individual’s location privacy. Swanlund et al.~\cite{swanlund_street_2020} propose a new method for masking locations of individuals. Instead of displacing locations randomly to other houses (which is done in traditional methods), they move them along a street in the OSM road network. One of the advantages of this method is that OSM data is readily available. For other methods address data and/or population data are required, both of which are more difficult to get hold of. 


\section{Conclusions and Future Work}
\label{section:conclusions}
OSM is mostly used as a basemap, whether for dashboards or for other services. This is not unique to OSM, as there are many dashboards and services using Google Maps and the like. The only cases we have found (so far) that actually used OSM data was when the restrictions called for complex spatial queries that could be completed by OSM-based tools. Yet, this is still not something that is inherently unique to OSM and could be developed also with other frameworks. Given the unique nature of OSM we had expected a more widespread utilisation in the situation of a global pandemic. However, OSM usage is still impressively high and global. Humanitarian efforts in OSM are well suited for disasters and hazards, but not to rolling events like epidemics. While OSM has a very flexible data model and many easy-to-use data contribution methods there is a tendency to map permanent entities (at least for the short term) while much of the mapping during COVID-19 is all about temporary response (e.g. stocks of masks, exposure, supermarket queuing, changes in opening times). There is probably a need to produce a practice of COVID-19 tagging in OSM which requires discussion and coordination in order to make tagging practices fit for pandemic response purposes. Given the global nature of the pandemic, such non-trivial efforts should in principle involve all OSM communities worldwide and would benefit from the coordination of the OpenStreetMap Foundation, which supports the OSM project but does not take decisions about tags~\cite{osmf}. However, the establishment of global COVID-19 tagging practices appears to be hard, not only because of the traditional differences in OSM tagging practices across the world~\cite{davidovic2016tagging}, but also due to the legal and ethical aspects that COVID-19 information might bring, at least in some countries or areas of the world.

OSM provides citizens and agencies an easy way to contribute or help in a pandemic response as opposed to not being able to contribute to authoritative datasets. Even if OSM data does not exist before the event it can be created almost in real-time by citizens locally or around the world. The data is accurate and high quality. The increased tile usage in badly affected COVID-19 areas also suggests that there is a need for freely accessible map services. OSM has been utilised in a wide variety of ways during the early phase of the pandemic. However, because OSM is truly open data we may never know of all the uses of OSM data during this period and estimation of usage could be difficult.

There are a number of very interesting questions for future work. Further investigation is required to understand if, and why, governments predominantly used proprietary tools and basemaps during the COVID-19 pandemic whilst research institutions, universities, community organisations used OSM. Analysis of the OSM contribution history will help to understand COVID-19 related OSM data contributions and/or data contributions during the pandemic. This could provide insights into possible correlations between the volume/activity/nature of contributions and the spread/evolution of the pandemic in a given country. The contribution of new data in OSM to address pandemics such as COVID-19 followed different contribution patterns than those observed before and these will offer fruitful grounds for future research work~\cite{editorial2020}. In Mooney and Juhász ~\cite{doi:10.1177/2043820620934926} the authors comment that many web-based maps produced during the early stages of the COVID-19 pandemic appear different or even contradictory. Urgent attention is required in order to consider how to deliver this information effectively within the constraints of the web-based map.
 Finally, it would be very interesting to consider a deeper exploration of the causes for differences between OSM communities during the pandemic given that COVID-19 has affected both developed and developing regions in the world.

\bibliographystyle{unsrt}  
\bibliography{mooney-et-al-bib.bib}  





\end{document}